# He and Ne Ages of Large Presolar Silicon Carbide Grains: Solving the Recoil Problem

*Ulrich Ott*[A], *Philipp R. Heck*[B,C], *Frank Gyngard*[D], *Rainer Wieler*[B], *Frédéric Wrobel*[E], *Sachiko Amari*[D] *& Ernst. Zinner*[D]

[A]Max-Planck-Institute for Chemistry, Joh.-J.-Becher-Weg 27, D-55128 Mainz, Germany;

[B]Institute of Isotope Geology and Mineral Resources, ETH Zürich, CH-8092 Zürich, Switzerland;

[C]Present address: Department of Geology and Geophysics, University of Wisconsin, Madison, WI 53706, USA;

[D]Laboratory for Space Sciences and the Physics Department, Washington University, St. Louis, MO 63130, USA;

[E]Institut d'Electronique du Sud, Université de Montpellier 2, F-34095 Montpellier, France.
E-mail: ott@mpch-mainz.mpg.de

**Abstract.** Knowledge about the age of presolar grains provides important insights into Galactic chemical evolution and the dynamics of grain formation and destruction processes in the Galaxy. Determination via the abundance of cosmic ray interaction products is straightforward, but in the past has suffered from uncertainties in correcting for recoil losses of spallation products. The problem is less serious in a class of large (tens of μm) grains. We describe the correction procedure and summarize results for He and Ne ages of presolar SiC "Jumbo" grains that range from close to zero to ~ 850 Myr, with the majority being less than 200 Myr. We also discuss the possibility of extending our approach to the majority of smaller SiC grains and explore possible contributions from trapping of cosmic rays.

# 1 Introduction

Primitive meteorites contain microscopic grains of stardust, which survived from times before the Solar System was born (Clayton & Nittler 2004; Lodders & Amari 2005; Zinner 2007). Studying these grains, which originated from a variety of stellar sources, is crucial to our understanding of the formation of elements in stars, dust destruction processes in the interstellar medium (ISM), and processes during formation of our Solar System. Silicon carbide (SiC) is the most widely studied type of presolar dust in meteorites. Most SiC grains originated in the outflows of Asymptotic Giant Branch (AGB) stars, i.e. solar-like stars in their final stages (e.g., Zinner 2007). They are the main carrier of Ne-G (Tang & Anders 1988a), a noble gas component produced by nucleosynthesis during He shell burning in AGB stars (Gallino et al. 1990).

For understanding processes in the ISM, a reliable determination of the period between formation of the grains and their incorporation into the early solar system is important. Conventional radiometric dating is hindered by small grain size, as well as the anomalous isotopic composition of essentially every element (Meyer & Zinner 2006). An alternative approach is to determine the length of time the grains were exposed to Galactic cosmic rays (GCR), by measuring GCR-produced nuclides, in particular, rare noble gas isotopes (Tang & Anders 1988b; Lewis, Amari & Anders 1994; Ott & Begemann 2000; Ott et al. 2005). Studies made on aggregates of ~micron-sized presolar SiC grains ("bulk samples") found GCR exposure ages of roughly $10^8$ years or less, considerably shorter than estimated lifetimes of interstellar dust (~$5 \times 10^8$ years; Jones et al. 1997). The first results (Tang & Anders 1988b; Lewis et al. 1994), however, were invalidated when it was realized that recoil losses of GCR-Ne from micron-sized grains are much larger than assumed (Ott & Begemann 2000). Apparently much of what had been assumed to be cosmogenic Ne must have had, in fact, a nucleosynthetic origin. Short GCR exposure ages of less than a few $10^7$ years were implied by spallation Xe, which is much less affected by recoil loss (Ott et al. 2005).

Gyngard et al. (2007, 2008; and this volume) recently reported longer presolar ages of between $4 \times 10^7$ and $1 \times 10^9$ years for very large (5-60 μm) individual presolar SiC grains based on measured excesses of $^6$Li. For such large grains, recoil losses of Ne are less of a problem, and even a sizeable fraction of cosmogenic He may have been retained. Thus, one can confidently also determine their He and Ne presolar exposure ages, since the identification of GCR-produced $^{21}$Ne and $^3$He is quite straightforward and production systematics are relatively well understood (Heck et al. 2008a, b). While a complete discussion of the results is given by Heck et al. (2008b), we here report the basic data but focus primarily on details of the recoil loss correction and provide only a short summary and updated discussion of exposure ages. We furthermore explore whether extending the present approach to smaller grains is possible and whether trapping of – in addition to production by - cosmic rays may have made a noticeable contribution.

## 2  Cosmogenic He and Ne in large SiC grains

### 2.1  LS and LU grains

The large silicon carbide grains analyzed here were from the LS and LU series isolated by Amari, Lewis, & Anders (1994) from the Murchison meteorite using a combination of chemical and physical separation steps. LS+LU grains are quite unique in size, shape and isotopic patterns (Amari et. al. 1994; Virag et al. 1992; Gyngard et al. this volume). A further characteristic is their low content of trace elements, including the noble gases, in comparison with other populations of presolar SiC. Carbon and Si isotopic compositions of the grains analyzed for He and Ne are given by Heck et al. (2008b). Most are of the mainstream type, while three grains (L2-12, L2-27, L2-57; Table 1) are of type AB ($^{12}$C/$^{13}$C < 10; e.g., Zinner 2007). L2-25 with $^{12}$C/$^{13}$C = 11.9 may also be of type AB.

## 2.2 Helium and Neon results

Helium and Ne data were obtained at the ETH Zürich with a high-sensitivity noble gas mass spectrometer with compressor source. For gas extraction the grains were bombarded by a Nd-YAG laser (Heck et al. 2007, 2008b). Isotopic data for 19 grains with detectable cosmogenic Ne are displayed in Figure 1. Not included are two grains where $^{22}$Ne concentrations – and consequently the plotted ratios– have very large uncertainty, but where nevertheless an upper limit (L2-08) and a value with large error (L2-16) for the $^{21}$Ne exposure age could be derived. Table 1 lists the concentrations of cosmogenic $^{21}$Ne and $^{3}$He together with the recoil corrections and the inferred presolar ages. For calculating the abundance of cosmogenic Ne, the data were treated as a 3-component mixture of: a) trapped Ne-G from the He shell of an AGB star of 1.5 solar masses ($^{21}$Ne/$^{22}$Ne = 5.9x10$^{-4}$, Gallino et al. 1990; $^{20}$Ne/$^{22}$Ne = 6.5x10$^{-2}$, Heck et al. 2007); b) interstellar cosmogenic Ne in SiC ($^{21}$Ne/$^{22}$Ne = 0.574, $^{20}$Ne/$^{22}$Ne = 0.735; Reedy 1989); and c) "trapped Ne", mostly from the extraction blank having the composition of air. For grain L2-03 ($^{21}$Ne/$^{22}$Ne = 0.22, $^{20}$Ne/$^{22}$Ne = 11.8), solar Ne instead of air was used as the third component. For three grains with no detectable cosmogenic $^{21}$Ne - where no ages have been given in Heck et al. (2008b) - generous two-sigma upper limits based on the total abundance of $^{21}$Ne are listed in Table 1, while for two more grains this approach yielded more stringent upper limits than those given by Heck et al. (2008b). Results for these five grains are given in italics and parentheses. A completely cosmogenic origin was adopted for $^{3}$He (Heck et al. 2008b; but see also Sec. 5.2).

**Table 1.** Cosmogenic $^{21}$Ne and $^{3}$He, recoil retention and recoil-corrected presolar ages of large SiC grains. The retention percentage combines direct production of $^{3}$He with that via $^{3}$H. Upper limits are in italics, upper limits based on total $^{21}$Ne are given in italics and parentheses.

| Grain | Size [µm] | | $^{21}$Ne$_{cos}$ [10$^{-8}$ cc/g] | $^{21}$Ne ret. [%] | $^{21}$Ne age [Myr] | $^{3}$He$_{cos}$ [10$^{-8}$ cc/g] | $^{3}$He ret. [%] | $^{3}$He age [Myr] |
|---|---|---|---|---|---|---|---|---|
| L2-01 | 7.3 | | 16.7 | 50.7 | 56 | 207 | 18.3 | 271 |
|  | | ± | 8.4 | | 30 | 11 | | 14 |
| L2-03 | 35.6 | | 218 | 88.5 | 439 | 43.7 | 33.7 | 30 |
|  | | ± | 24 | | 47 | 1.8 | | 1 |

| | | | | | | | |
|---|---|---|---|---|---|---|---|
| L2-04 | 9.2 | ± | 20 27 | 58.8 | < 224 | < 50 | 20.3 | < 58 |
| L2-05 | 5.7 | ± | (~ 0) | 42.2 | (~ 0) | 87 54 | 16.3 | 128 80 |
| L2-06 | 17.3 | ± | 7.0 0.5 | 76.7 | 15 1 | 20.5 1.4 | 26.2 | 18 1 |
| L2-07 | 8.8 | ± | (< 29) | 57.2 | (< 89) | 10.2 9.4 | 19.9 | 11 11 |
| L2-08 | 18 | ± | 0.3 0.8 | 77.5 | < 3 | 25.9 6.0 | 26.6 | 22 5 |
| L2-09 | 10.3 | ± | 12.2 3.0 | 62.5 | 33 8 | 21 14 | 21.3 | 23 16 |
| L2-10 | 9 | ± | 7.5 10.3 | 58.0 | < 83 | 65 19 | 20.1 | 77 23 |
| L2-11 | 11.8 | ± | 45.4 10.5 | 66.7 | 119 28 | 60 15 | 22.6 | 63 16 |
| L2-12 | 11 | ± | 48.2 7.0 | 64.6 | 131 20 | 287 31 | 21.9 | 314 34 |
| L2-13 | 7.8 | ± | (<45) | 53.0 | (< 89) | 88 21 | 18.9 | 111 27 |
| L2-14 | 11 | ± | 35.1 6.4 | 64.6 | 95 19 | 164 24 | 21.9 | 179 26 |
| L2-15 | 9.6 | ± | (< 46) | 60.2 | (< 135) | 88 35 | 20.7 | 101 41 |
| L2-16 | 8.1 | ± | 29 27 | 54.4 | 93 88 | < 20 | 19.2 | < 24 |
| L2-17 | 8.4 | ± | (<8) | 55.6 | (< 25) | 35 11 | 19.5 | 42 14 |
| L2-18 | 15.6 | ± | 5.7 2.3 | 74.3 | 11 5 | 29.5 5.1 | 25.2 | 27 5 |
| L2-19 | 9.3 | ± | 42.7 9.2 | 59.1 | 126 27 | 87 12 | 20.4 | 102 14 |
| L2-25 | 4.9 | ± | 180.0 77.6 | 37.3 | 854 372 | 166 57 | 15.1 | 264 91 |
| L2-27 | 2 | ± | 1170 657 | 16.0 | -- | < 987 | 9.0 | < 2639 |
| L2-57 | 5.8 | ± | 135.0 30.7 | 42.7 | 558 129 | 123 43 | 16.4 | 179 63 |

# 3 Production and Recoil

## 3.1 Production rates

Production rates used for calculating the exposure ages are based on the estimates of Reedy (1989) for proton-induced reactions on Si and C, which are multiplied by a factor 1.33 to take into account production by α-particles (Ott et al. 2005). The resulting production rates are 4.15x10$^{-8}$ cc/g per Myr for $^3$He, and 5.60x10$^{-9}$ cc/g per Myr for $^{21}$Ne. For a discussion of the uncertainties see Heck et al. (2008b).

## 3.2 Recoil effects

*Historical Notes.* As described in the introduction, accounting for recoil losses from ~ μm-sized presolar SiC grains has been a major problem, and the early results reported by Tang & Anders (1988b) and Lewis et al. (1994) became invalid when Ott & Begemann (2000) found in irradiation experiments that recoil losses were far more extensive than assumed by the previous authors. Several facts had conspired that led to these underestimates. A simple error was introduced by Tang & Anders (1988b) in scaling the range (see Fig. 3 in Ott & Begemann 2000). More serious are flaws in the interpretation by Ray & Völk (1983) - on which Tang & Anders (1988b) relied - of the momentum distribution of products from the fragmentation of high-energy C and O projectiles given by Greiner et al. (1975). Morissey (1989), in a survey of a large number of relevant experimental results, obtained an empirical relationship (see Fig. 3 there) between average momentum and the square root of the mass difference between target and product. Reassuringly, this relationship, combined with range-energy relations from the SRIM code (Ziegler 2004), results in an average recoil range of ~2.2 μm for $^{21}$Ne produced in SiC, quite similar to the ~2.5 μm inferred by Ott & Begemann (2000). The Greiner et al. (1975) data, which were obtained for light products (A<15), on the other hand, fall below the correlation line and hence their applicability to the case of $^{21}$Ne seems questionable at best.

More serious even than the *application* of the Greiner et al. (1975) data set to the case of $^{21}$Ne is the *handling* of these data by Ray and Völk (1983). This is because these authors simply used the *average momentum* given by Greiner et al. (1975), which includes *directional* averaging and which for this reason is close to zero in the frame of the moving C and O nuclei (equivalent to the target elements of SiC in our case). However, a nucleus with sufficient momentum will be lost due to recoil even if - in calculating the average - its momentum is largely canceled by product nuclei emitted in other directions. For a correct description of recoil losses the momentum *distribution* needs to be converted into an energy distribution, which then is folded with range-energy relations. We have done so for the case of $^3$He (see below), but not for $^{21}$Ne.

*Neon recoil*. In Heck et al. (2008b), we have used the average recoil range of ~2.5 μm for $^{21}$Ne in SiC as inferred by Ott & Begemann (2000) from the losses observed in their artificially irradiated SiC samples. We approximated the grains as spherical and applied the corresponding geometrical relationship described in e.g., Tang & Anders (1988b) and Ott & Begemann (2000). Here, instead of a constant recoil range, we use a distribution based on theoretically derived energy spectra for $^{21}$Ne (Fig. 2) calculated as described in Wrobel (2008). Retention values for the individual grains are listed in Table 1. In the size range of interest the results from both approaches are virtually identical (Fig. 3).

*Helium recoil*. Recoil retention of $^3$He is based on the Greiner et al. (1975) momentum distributions of $^3$He and $^3$H (assumed to contribute half of the final $^3$He yield) nuclei in the fragmentation of C and O nuclei. Since there is no significant difference between fragmentation of C and O and also no significant difference between the momentum distribution produced from C at the two energies (1.05 and 2.10 GeV/n) employed in the Greiner et al. (1975) experiments, we used the average parameters characterizing their three Gaussian distributions. After multiplying by √3 (to include the momentum perpendicular to the beam axis in the experiment; see eq. 6 in Morissey 1989), these were converted into energy spectra for $^3$He and $^3$H (the one for $^3$He is shown in Fig. 2). The energy spectra in turn were folded with range-energy relations (SRIM Code; Ziegler 2004) and the formalism describing retention by spherical grains (Tang & Anders 1988b; Ott & Begemann 2000) to derive retention values as a function of grain size (Fig. 4). Note that while the momentum and energy spectra for directly produced $^3$He and tritium are virtually the same, due to the fact that tritium carries only one nuclear charge, its range is ~4x longer than that of $^3$He over most of the energy range, and losses are accordingly higher.

## 4 Helium and Neon Ages

Recoil-corrected presolar $^{21}$Ne and $^3$He ages are listed in Table 1, where the ~ 1 Myr recent exposure in the Murchison meteorite has also been taken into account. We concentrate on the $^{21}$Ne ages, which we calculated based on the Wrobel energy spectra (Fig. 2) for recoil correction rather than a fixed range of 2.5 μm as was done by Heck et al. (2008b). Results differ only slightly as noted above (cf. Fig. 3). The largest effect is for L2-25, whose size is only 4.9 μm and in which expected retention is 39% instead of 31%, resulting in a reduction in age from 1060 to ~850 Myr. In all other cases, the difference is much smaller. For a discussion of the $^3$He ages and a comparison between He and Ne ages, we refer to Heck et al. (2008b).

The resulting ages are shown in Figure 5. The most notable observation is that most ages are short (< 200 Myr), clearly shorter than expected lifetimes of presolar grains of ~ 500 Myr (Jones et al. 1997). Only one of the mainstream SiC grains analyzed here falls into that age range. A possible explanation for young ages based on a Galactic Merger 1.5-2 Gyr before Solar System formation (Clayton 2003) is discussed in Heck et al. (2008b) and Ott et al. (2005). Interestingly, the two grains with the longest $^{21}$Ne exposure are of type AB or possibly type AB (grains L2-57 and L2-25, respectively). As for the other AB grains, L2-12 has an unremarkable age (131 Myr), while the smallest analyzed grain (L2-27, 2 μm; not plotted) may also have a very long exposure, which is, however, zero within 2 sigma. Since the AB grains are among the smallest analyzed here, it is not clear whether the difference is related to origin (AB vs. mainstream), grain size, or simply poor statistics (see Heck et al. 2008b, for further discussion). The latter also may or may not be an explanation for the fact that Li ages - which were obtained on a less diverse set of grains - seem to be considerably longer. Obviously, it will be necessary to determine Li and He/Ne ages on the same grains.

## 5  Additional Considerations

### 5.1  The Case of the Small Grains

A better understanding of the recoil correction should – in principle – allow us to extend the age dating to the smaller grain sizes, which are more typical of presolar SiC than the large grains from the LS+LU series. There is, however, an important difference. For the LS+LU grains analyzed here, cosmogenic Ne is a dominant component and the inferred abundances (and thus the ages) depend only weakly on the choice of the $^{21}Ne/^{22}Ne$ ratio in the G component. We have used $(^{21}Ne/^{22}Ne)_G = 5.9 \times 10^{-4}$ (Gallino et al. 1990), but a higher value such as $3.3 \times 10^{-3}$, predicted by Karakas et al. (2008) for a 3 $M_\odot$ solar-metallicity AGB star with the upper experimental limit to the $^{18}F(\alpha,p)$ reaction rate, would change the inferred cosmogenic $^{21}Ne$ by less than 5 % in most cases. On the other hand, for all of the Murchison K series separates, the extrapolated $^{21}Ne/^{22}Ne$ ratio of the G and the cosmogenic component *combined* is $< 2 \times 10^{-3}$, i.e. lower than the upper limit to the ratio in the G component alone (Lewis et al. 1994). Obviously, in this situation it is not possible to draw any useful conclusions about the cosmogenic component, unless there is a better understanding of the nucleosynthetic component. As noted by Ott & Begemann (2000), there is a correlation in the Lewis et al. (1994) data between $(^{21}Ne/^{22}Ne)$ of the combined G and cosmogenic components and the $^{86}Kr/^{82}Kr$ ratio of the G-component that is sensitive to the details of s-process nucleosynthesis. This may allow, once a thorough understanding of AGB nucleosynthesis has been achieved, to cross-calibrate the two ratios and use simultaneously measured Kr to infer the Ne isotopic composition of the G component.

### 5.2  Trapping versus Production?

Galactic cosmic rays not only produce new nuclides in presolar grains, but if sufficiently slowed down, can be implanted and thus become trapped. Ott & Huss (2008) have suggested GCR trapping to explain the He isotopic signatures in presolar diamond. Because of the nanometer-size of the diamonds, their model requires slowing down the GCR to very low energy in an ambient medium. In the case of grains that are tens of μm in size, the low-energy part of the cosmic rays can be both slowed down and efficiently trapped. As pointed out by Ott & Huss (2008); such a component should primarily show up in $^3$He, and we attempt to obtain a crude estimate of the possible contribution to this nuclide. Our estimate is based on the proton flux (average of several spectra) in the ISM used by Reedy (1989) to derive production rates (Sec. 3.1), which in the (poorly known) low-energy range 0-3 MeV is ~ 0.10 protons cm$^{-2}$ s$^{-1}$ (Reedy, pers. comm.). With a $^3$He/proton ratio of ~ 0.02 as measured at higher energies (references in Ott & Huss 2008), the corresponding $^3$He flux is ~2x10$^{-3}$ cm$^{-2}$ s$^{-1}$ in the energy range 0-9 MeV. Such $^3$He nuclei have a range in SiC of less than 50 μm according to the SRIM code (Ziegler 2004). Assuming that – for reasons of geometry and energy distribution – a spherical ~ 50 μm grain traps about half of these $^3$He ions that it encounters, the resulting $^3$He GCR *trapping* rate is on the order of ~2x10$^{-8}$ cc/(g Ma). This is roughly half the GCR *production* rate that we used for calculating the exposure ages (Sec. 3.1). In other words, for grains in the size range of some tens of microns, trapping of cosmic ray $^3$He may make a noticeable contribution to "cosmogenic" $^3$He, resulting in (somewhat) smaller $^3$He ages as reported here and in Heck et al. (2008b). Whether trapping may be more or less important for smaller grains, depends on the detailed energy spectrum, which is poorly known. On the other hand, since in the cosmic rays $^{21}$Ne/$^3$He is ~ 1/300, any trapping contribution to $^{21}$Ne would obviously be insignificant. Better knowledge of the flux and composition of the low energy part of cosmic rays will be essential to better constrain the effects of trapping.

## 6  Summary

Cosmic ray exposure ages of presolar silicon carbide grains in the size range ~5 to ~ 40 μm, determined in this study and by Heck et al. (2008b), are mostly less than 200 Myr, i.e. shorter than expected lifetimes of interstellar grains. They are also shorter on average than CRE ages determined from $^6$Li excesses on another ensemble of grains. Determining He/Ne and Li ages on the same grains is an important future task. Grains of type AB seem to be older than mainstream SiC grains, but the observed difference may be due to a grain size effect or poor statistics.

Recoil loss corrections for grains in the range studied here can be reliably performed. Extension to smaller grain sizes, where – due to higher contents of Ne-G – cosmogenic $^{21}$Ne is less prominent requires reliable knowledge of the Ne-G isotopic composition. Its determination should be a primary task for future studies of AGB star nucleosynthesis. We have also considered the potential contribution of trapped cosmic rays to the observed "cosmogenic" He and Ne. While a significant trapping contribution to $^3$He appears possible, any contribution to $^{21}$Ne must be minor.

**Acknowledgements**

We thank Roy Lewis for providing the LS+LU grains, and H. Baur and M.M.M. Meier for technical and analytical support at ETH. This work has been supported by the Swiss National Science Foundation and by NASA Grant NNX08AG71G (EZ, PI).

**Figure Captions**

**Figure 1** Isotopic ratio $^{20}$Ne/$^{22}$Ne plotted vs. $^{21}$Ne/$^{22}$Ne in large presolar SiC grains. Errors are 1 sigma.

**Figure 2** Energy spectra (binned in 0.5 MeV steps; the sum totals 1 in each case) of cosmogenic $^{21}$Ne and $^{3}$He as used for predicting retention of these two nuclides. Data for $^{21}$Ne are based on calculations for $^{21}$Ne production on Si by 200 MeV protons according to the method of Wrobel (2008); those for (directly produced) $^{3}$He are based on the momentum distribution derived by Greiner et al. (1975) for fragmentation of high energy $^{12}$C and $^{16}$O projectiles as described in the text. The spectrum for $^{3}$H, which after decay contributes ~50% of the final $^{3}$He yield, is similar to that for directly produced $^{3}$He.

**Figure 3** Retention of spallation $^{21}$Ne in spherical SiC grains for the energy spectrum of $^{21}$Ne based on calculations as described by Wrobel (2008) and shown in Fig. 2; results for a constant recoil range of 2.5 µm (Ott & Begemann 2000) are shown for comparison. For grain sizes larger than 5 µm, i.e. in the range of interest here, they are virtually identical.

**Figure 4** Retention of spallation $^{3}$He in spherical SiC grains as a function of grain size, for energy spectra based on the momentum distribution given by Greiner et al. (1975) (see text and Figure 2). Retention for total $^{3}$He (= sum) is calculated assuming a 50% contribution by directly produced $^{3}$He and 50% contribution by tritium as a precursor.

**Figure 5** Ne ages (circles) are compared to the Li ages (crosses) of Gyngard et al. (2008; and this volume). Grey circles (Ne upper limit) depict upper limits based on upper limits to *cosmogenic* $^{21}$Ne, while open circles (Ne not detected) show upper limits based on upper limits to *total* $^{21}$Ne.

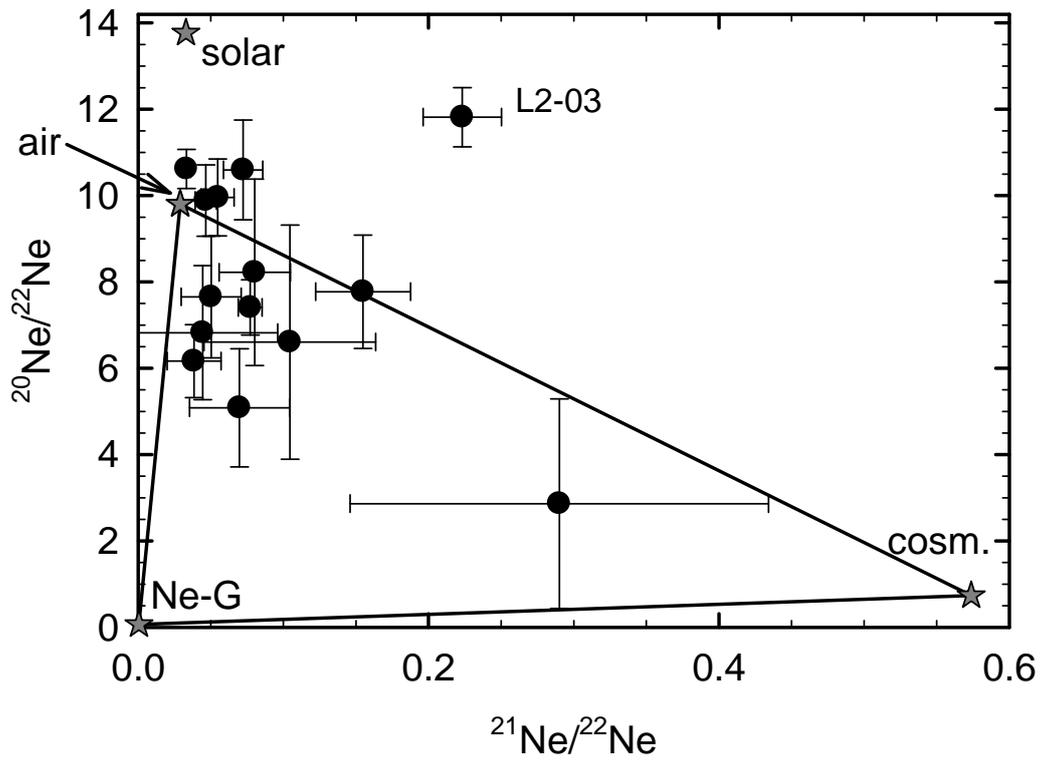

Fig. 1

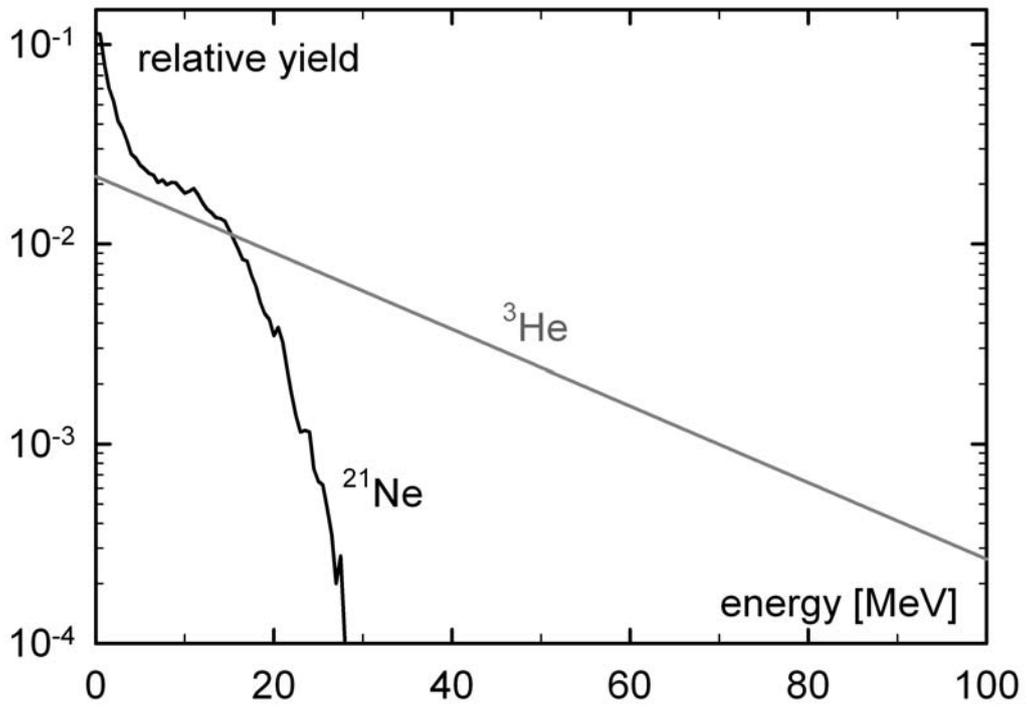

Fig. 2

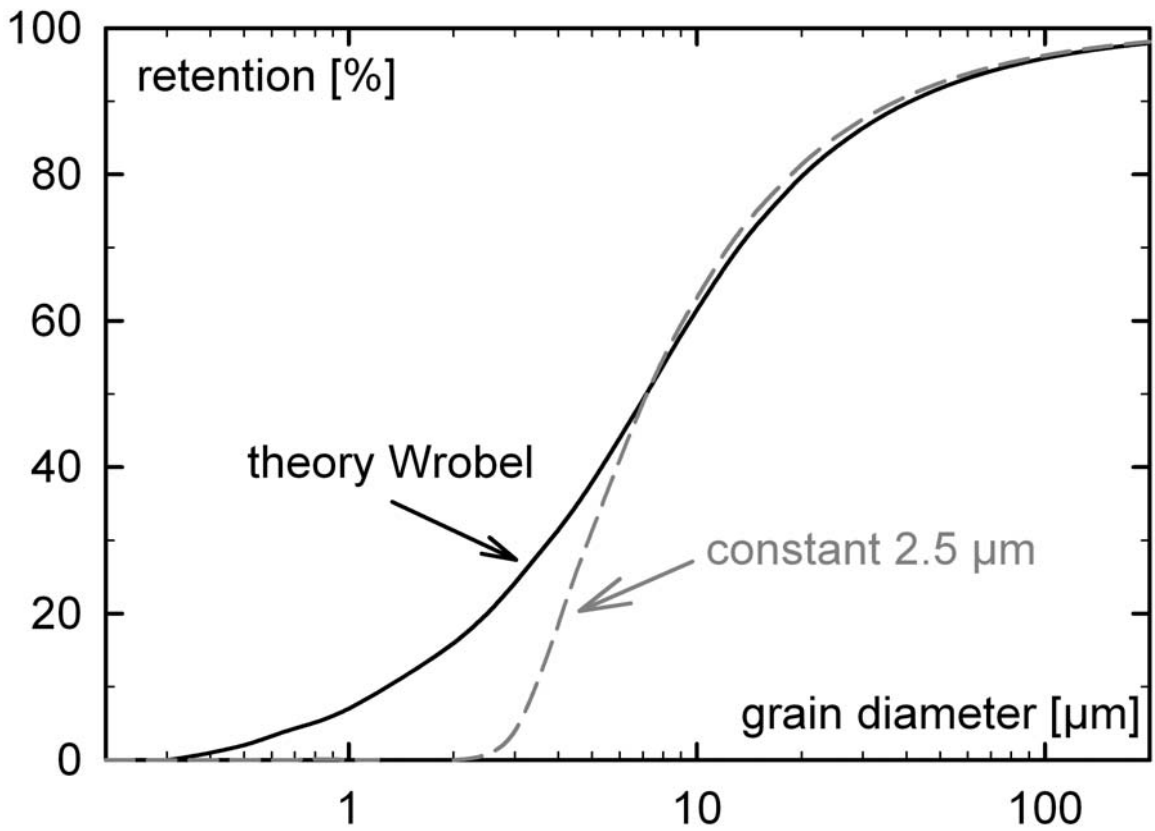

Fig. 3

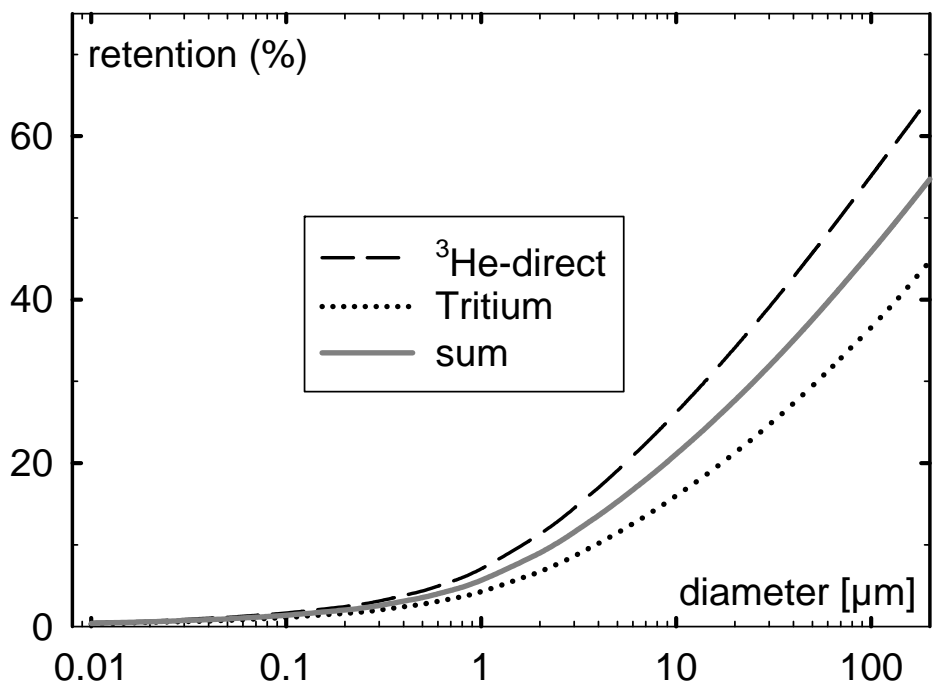

Fig. 4

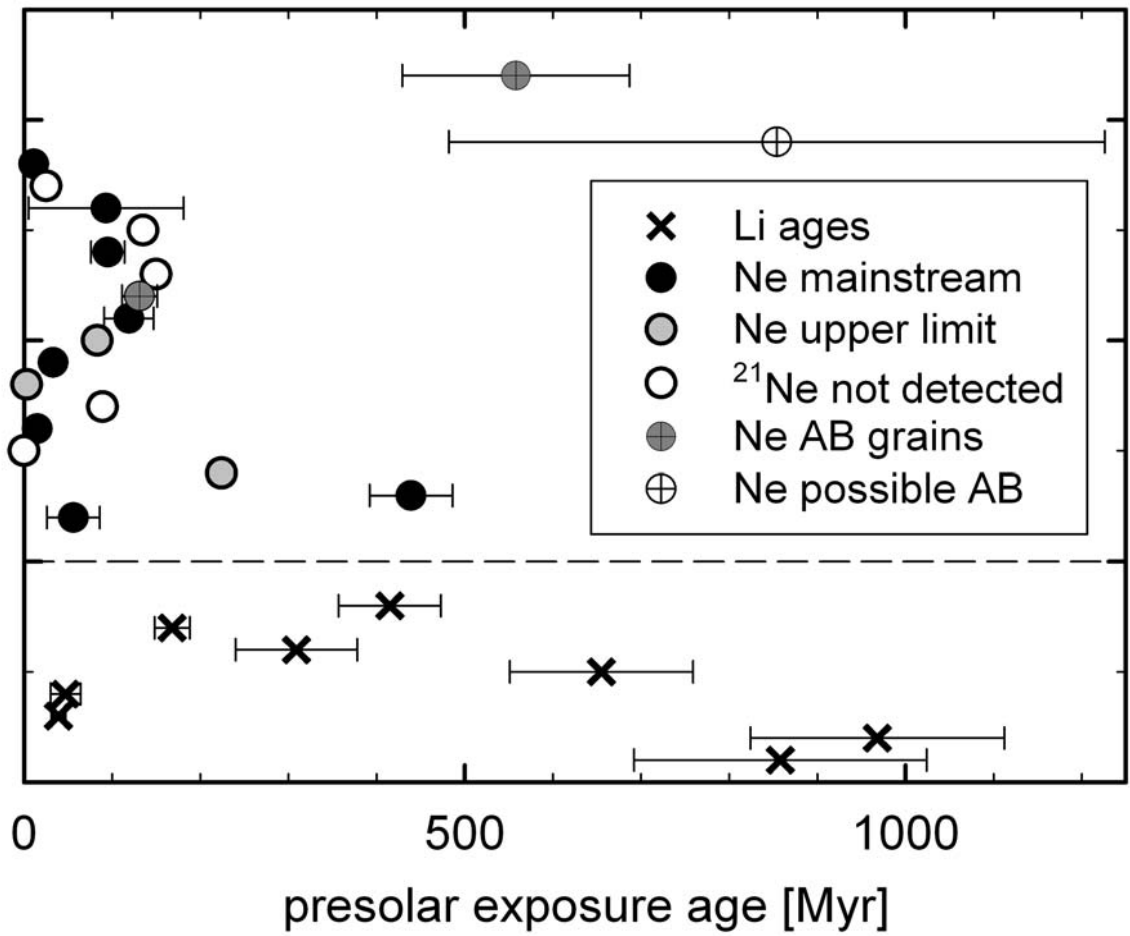

Fig. 5